\documentclass[apj]{emulateapj}

\usepackage{rotating}
\usepackage{epsfig}

\newcommand{\lt}{\ifmmode\,<\,\else \,$<$\,\fi}
\newcommand{\kms}{\ifmmode\,{\rm km}\,{\rm s}^{-1}\else km$\,$s$^{-1}$\fi}

\newcommand{\magarc}{\ifmmode {{{{\rm mag}~{\rm arcsec}}^{-2}}}
             \else {{{mag}$~${arcsec}$^{-2}$}}
             \fi}

\newcommand{\Lya}{Ly$\alpha$}

\shorttitle{On The Escape Fraction of Ionizing Photons in $z\simeq$4 Galaxies}

\begin{document}

\title{Keck Spectroscopy of Gravitationally Lensed $z\simeq4$ Galaxies:  Improved Constraints on the Escape Fraction of Ionizing Photons}


\author{Tucker A Jones\altaffilmark{1},
Richard S Ellis\altaffilmark{2}, 
Matthew A Schenker\altaffilmark{2},
Daniel P Stark\altaffilmark{3,4}
}

\altaffiltext{1} {Department of Physics, University of California, Santa Barbara, CA  93106}
\altaffiltext{2}{Department of Astrophysics, California Institute of 
                Technology, MS 249-17, Pasadena, CA 91125}
\altaffiltext{3}{Department of Astronomy and Steward Observatory, University of Arizona, Tucson AZ 85721}
\altaffiltext{4} {Hubble Fellow}

%\date{\today}

\begin{abstract}
The fraction of ionizing photons that escape from young star-forming galaxies is one of the largest uncertainties in determining the role of galaxies in cosmic reionization. Yet traditional techniques for measuring this fraction are inapplicable at the redshifts of interest due to foreground screening by the Lyman $\alpha$ forest. In an earlier study, we demonstrated a reduction in the equivalent width of low-ionization absorption lines in composite spectra of Lyman break galaxies at $z\simeq4$ compared to similar measures at $z\simeq3$. This might imply a lower covering fraction of neutral gas and hence an increase with redshift in the escape fraction of ionizing photons. However, our spectral resolution was inadequate to differentiate between several alternative explanations, including changes with redshift in the outflow kinematics. Here we present higher quality spectra of 3 gravitationally lensed Lyman break galaxies at $z\simeq4$ with a spectral resolution sufficient to break this degeneracy of interpretation. We present a method for deriving the covering fraction of low-ionization gas as a function of outflow velocity and compare the results with similar quality data taken for galaxies at lower redshift. We find a significant trend of lower covering fractions of low-ionization gas for galaxies with strong  \Lya\ emission. In combination with the demographic trends of \Lya\ emission with redshift from our earlier work, our results provide new evidence for a reduction in the average H {\sc i} covering fraction, and hence an increase in the escape fraction of ionizing radiation from Lyman break galaxies, with redshift.

\end{abstract}


\section{Introduction}\label{sec:intro}

Star forming galaxies are the leading candidate for the source of ultraviolet photons required to reionize the universe. Several lines of evidence indicate that reionization was underway by $z=11$ and ended a few hundred Myr later at $z\simeq6$--7 (e.g., \citealt{schenker2012,mortlock2011,hinshaw2012}). Deep near-IR imaging with the Hubble Space Telescope has provided good constraints on the UV luminosity density of star forming galaxies during the reionization epoch (e.g., \citealt{ellis2013,oesch2013}) which indicate that the ongoing star formation is likely capable of producing the required ionizing flux. However, it is unclear whether this radiation is able to escape from galaxies and actually ionize the intergalactic medium (IGM). Based on estimates of the total UV luminosity density and IGM clumping factor, the required escape fraction $f_{esc}$ of hydrogen-ionizing photons is $\gtrsim 0.2$ (e.g., \citealt{robertson2013}). The precise value of $f_{esc}$ is a key uncertainty in determining the role of galaxies in reionization.

Direct measurements of $f_{esc}$ during the reionization epoch are essentially impossible, not only because of the faint apparent luminosity, but also because foreground IGM attenuates the ionizing flux to undetectable levels even at $z\gtrsim4$. Direct imaging and composite spectra of galaxies at lower redshift has established a modest average $f_{esc} \simeq 0.05$ for Lyman break galaxies (LBGs) at $z=3$ \citep{bogosavljevic2010}. If star forming galaxies dominated reionization, $f_{esc}$ must have been higher at earlier times. This paper is concerned with improving the constraints on $f_{esc}$ at higher redshifts for which direct measurements are not practical. Our methodology is as follows: $f_{esc}$ is set by the areal covering fraction of hot stars by H {\sc i} such that $f_{esc} = 1 - f_c$, and $f_c$ can be inferred from intermediate dispersion spectroscopy of interstellar UV absorption lines.
The difficulty, of course, is that LBGs at redshift $z>3$ are very faint, so securing suitably high quality absorption line spectra of individual examples is a very challenging proposition. In an earlier paper \citep{jones2012} we therefore analyzed the average properties of LBGs at $z\simeq4$--5 derived from composite spectra in a manner similar to that pioneered at $z\simeq3$ by \citet{shapley2003}.

A particular motivation for the present study was the discovery of a marked reduction with increasing redshift in the equivalent width of low-ionization absorption lines at fixed UV luminosity, suggestive of changes in either the kinematic profile or the covering fraction of neutral gas \citep{jones2012}. Additionally, spectra of individual galaxies show a trend of increasing \Lya\ emission equivalent width with redshift which we argued could reflect evolution in the H~{\sc i} covering fraction \citep{stark2010}, especially given the strong correlation of \Lya\ and low-ionization absorption lines \citep{shapley2003,jones2012}.
A lower covering fraction for the neutral gas within typical LBGs would be particularly important as it could imply a higher escape fraction of ionizing photons. The composite spectra discussed by \citet{jones2012} did not have adequate resolution to distinguish the effects of reduced covering fraction and kinematics, and so the present paper takes this investigation one step further by attempting to resolve this important ambiguity. Here we present higher resolution spectra of 3 gravitationally lensed $z\simeq4$ LBGs. Although their unlensed luminosities are typical of the constituent galaxies comprising the \citet{jones2012} composite, their individual lensed magnitudes are much brighter enabling comparable signal to noise to the stack of LBGs discussed in our earlier paper.

Throughout the paper we adopt a flat $\Lambda$CDM cosmology with $\Omega_{\Lambda}=0.7$, $\Omega_{M}=0.3$, and H$_0 = 70$ \kms Mpc$^{-1}$. All magnitudes are in the AB system \citep{oke1974}.

\section{Gravitationally-Lensed $z\simeq4$ Galaxies}\label{sec:data}

The spectra comprising the composite published by \citet{jones2012} were taken with the 600 line mm$^{-1}$ DEIMOS grating with a resolution of $\simeq$3.5 \AA, although uncertain systemic redshifts led to reduced resolution of the composite spectrum (corresponding to a velocity resolution of $\sim 450$ \kms\ FWHM). The composite comprised galaxies with $z'_{AB}=24$--26 with 90\% completeness to $z'_{AB}=25$ at $z\simeq4$. The stacked spectrum reveals multiple low ionization lines such as Si {\sc ii} $\lambda$1260, O {\sc i} $\lambda$1302 + Si {\sc ii} $\lambda$1304, and C {\sc ii} $\lambda$1334 in the region where the signal/noise is optimal. As discussed by \citet{jones2012}, these lines are normally saturated so the line depth at a given velocity provides a measure of the areal covering fraction $f_c$ of O and B stars by neutral H {\sc i} gas along the line of sight. 

Typical LBGs are too faint for detailed line profile studies even at $z=2$, but strong gravitational lensing can boost the brightness of representative examples making such studies of individual sources a practical proposition. Studies of several lensed $z\simeq2$--3 LBGs have found that absorption velocities of low-ionization metal transitions range from $\sim-1000$ to +500 \kms\ with typical line centroids $v \sim -200$ \kms\ \citep{pettini2002,quider2009,quider2010,dessauges-zavadsky2010}. The mean low ionization absorption velocity in the \citet{jones2012} composite is similar, $v_{LIS} = -190$ \kms.

In a similar fashion, for the present analysis we have located 3 gravitationally-lensed LBGs at $z\simeq4$. Two are independent sources lensed by the well-studied cluster Abell 2390 and the third was located in the cluster J1621+0607 in the Sloan Digital Sky Survey. A2390\_H3 and H5 represent two distinct highly-elongated pairs of lensed images that were spectroscopically confirmed to be at different redshifts by, respectively, \citet{frye1998} and \citet{pello1999}. The tangential arc system in J1621+0607 was spectroscopically confirmed by \citet{bayliss2011}. The gravitational magnification factor is $\simeq$10 in each case \citep{pello1999}. We summarize the key properties in Table~\ref{tab:arcs}. The redshifts and absolute UV luminosities are representative of sources studied by \citet{jones2012}. In addition, for comparison purposes, we include an analysis of high quality spectra for 3 further lensed $z=2$--3 sources (courtesy of M. Pettini) in Table~\ref{tab:arcs}. These include the `Horseshoe' ($z$=2.38, \citealt{quider2010}), cB58 ($z$=2.73, \citealt{pettini2002}) and the `Cosmic Eye' ($z$=3.07, \citealt{quider2009}).

Spectra were taken with the 1200 line mm$^{-1}$ DEIMOS grating during two runs in October 2011 and June 2012. This provides a resolution of $\simeq$1.7 \AA\, corresponding to a velocity resolution of $\simeq$70 km s$^{-1}$, considerably better than for the composite discussed by \citet{jones2012}. Spectra of each galaxy covered wavelengths corresponding to at least 1175$-$1675 \AA\ in the rest frame.
The lensed sources in Abell 2390 were observed simultaneously with a multi-slit mask that sampled two images of each source. Seeing varied between 0\farcs4 and 1\farcs4 FWHM during the observations, and the bulk of the data used has seeing in the range 0\farcs7-0\farcs9. Some exposures ($\sim 10$\%) were affected by cirrus and are not included in the final addition. Total observing times for the final spectra are given in Table~\ref{tab:arcs}.

\begin{figure*}
\begin{center}
\includegraphics[width=\textwidth]{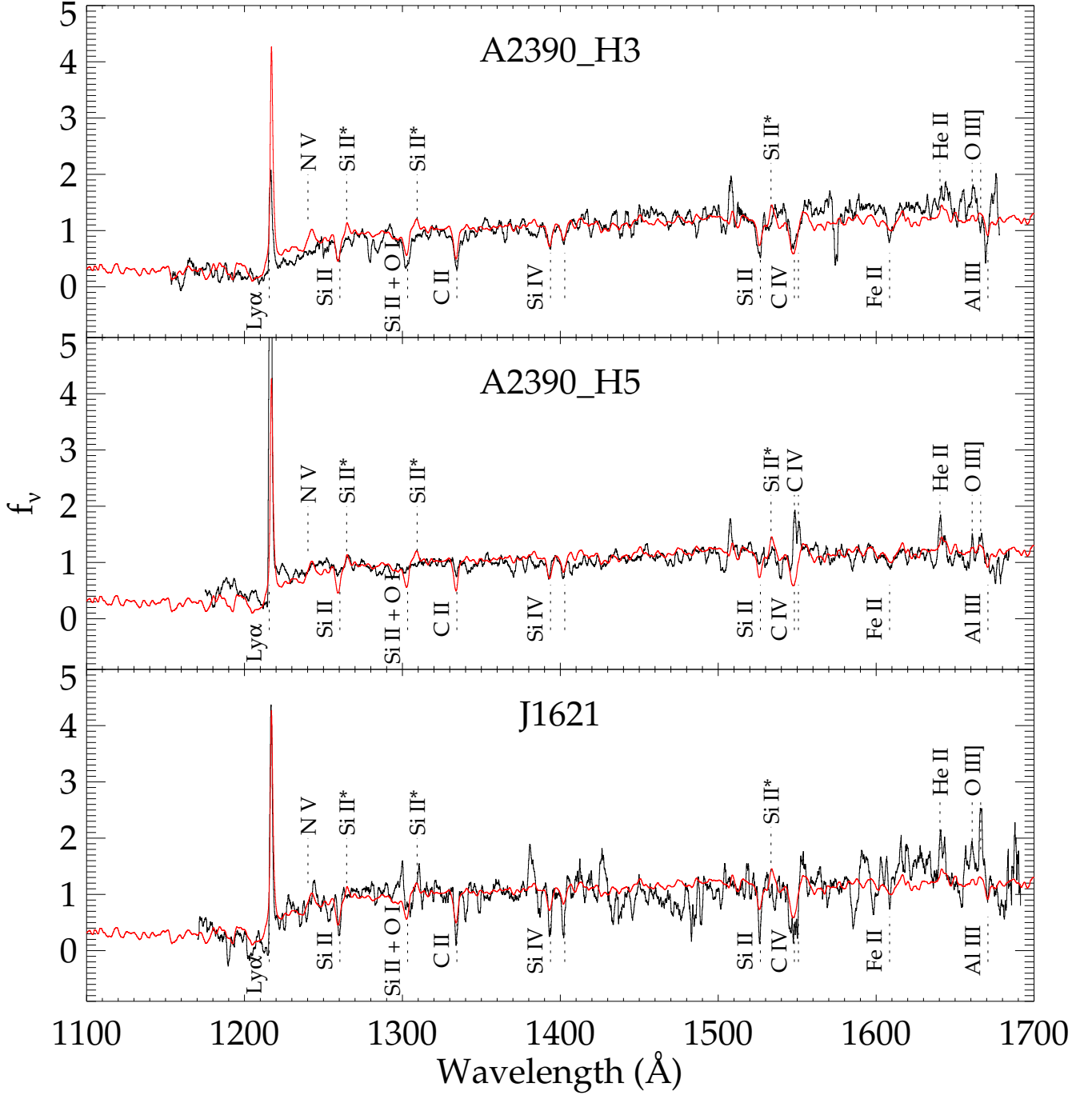}
\caption{
\label{fig:spectra}
Keck DEIMOS spectra of (top to bottom) A2390\_H3, A2390\_H5, and J1621. In each panel the black spectrum represents the lensed galaxy and the composite spectrum of 81 LBGs from the analysis of \citet{jones2012} is shown in red. All spectra are plotted in the rest frame with flux normalized such that median $f_{\nu}=1$ in the range $1250-1500$ \AA. Prominent spectral features are labeled.
}
\end{center}
\end{figure*}

The DEIMOS spectra were reduced and calibrated using the {\sc Spec2D} pipeline following the techniques discussed in detail by \citet{stark2010}. In the case of A2390\_H3, care was taken to ensure that the extracted spectrum was not contaminated by light from a nearby cluster member. Data from the October 2011 and June 2012 runs were reduced separately and the resulting one-dimensional spectra were combined with an inverse-variance weighted mean. Spectra of J1621 are affected by poor sky subtraction residuals, while the Abell 2390 arc spectra are of excellent quality. Spectra of different images of the Abell 2390 arcs were scaled to the same flux level before combining to a common wavelength scale with 0.7 \AA\, pixels, roughly Nyquist sampled. The final spectra, shown in Figure~\ref{fig:spectra}, reach an average continuum S/N per 70 \kms\, resolution element of 5 for J1621, 9 for A2390\_H3, and 10 for A2390\_H5 over the rest-frame wavelength range 1250--1650 \AA. This is comparable to that in the composite spectrum in \citet{jones2012} which has S/N equivalent to $\sim 10$ at the improved resolution of 70 \kms of our new data.

\section{Analysis}\label{sec:analysis}

\subsection{Systemic Redshift}\label{sec:redshift}

Accurate systemic redshifts are required in order to examine the kinematics of gas seen in absorption and the techniques for estimating these are discussed in detail in \citet{jones2012}. This is straightforward when nebular emission lines are visible, such as is the case in both J1621+0607 (O~{\sc iii}] $\lambda\lambda$1661,6) and A2390\_H5 (O {\sc iii}] $\lambda\lambda$1661,6, He {\sc ii} $\lambda$1640, C {\sc iv} $\lambda\lambda$1548,51). The strong emission from highly ionized species such as  He {\sc ii} and C {\sc iv} seen in A2390\_H5 is uncommon but has been observed in some high redshift starburst galaxies and signifies an extremely young, metal-poor, and hot stellar population (e.g. \citealt{fosbury2003,erb2010}). Alternatively they may signify the presence of an active galactic nucleus, but the narrow line widths ($50-185$ \kms\, FWHM, corrected for instrumental resolution) suggest an origin in star-forming H{\sc ii} regions.

No appropriate features are detected in the spectrum of A2390\_H3 and so we estimate the systemic redshift from low-ionization absorption lines using the method of \citet{jones2012}. This gives $z = 4.043 \pm 0.002 = z_{IS} + 190$ \kms, with uncertainty dominated by an rms difference $\sim 125$ \kms\, between redshifts derived from absorption lines and that obtained from nebular emission \citep{steidel2010}. The adopted systemic arc redshifts are listed in Table~\ref{tab:arcs}.

\subsection{Low-Ionization Covering Fraction}\label{sec:fcov}

Ideally we would measure the covering fraction of neutral hydrogen directly from spatially resolved H {\sc i} absorption. However, the only available transition (\Lya) is dominated by strong emission with net equivalent width $W_{\rm Ly\alpha} = 20-100$ \AA, and the observed line profile is complicated by resonant scattering in the extended circumgalactic medium (CGM; \citealt{steidel2011}). These effects are apparent from the \Lya\ line profiles which show redshifted emission as well as strong absorption arising from both the CGM and \Lya\ forest (Figure~\ref{fig:spectra}).
We therefore estimate the covering fraction of neutral hydrogen from absorption lines of heavier low-ionization species which arise in H {\sc i} gas, i.e., those with ionization potentials less than 1 Rydberg.

The covering fraction of any ion is related to its absorption line optical depth $\tau$ and residual intensity $I$ via
\begin{equation}\label{eq:fcov}
%\tau = - \ln{\left[ \frac{I - I_0(1-f_c)}{I_0 f_c} \right]}
\frac{I}{I_0} = 1 - f_c (1 - e^{-\tau})
\end{equation}
where $I_0$ is the continuum level. Optical depth is in turn related to column density as
\begin{equation}\label{eq:N}
\tau = f \lambda \frac{\pi e^2}{m_e c} N = f \lambda \frac{N}{3.768 \times 10^{14}}
\end{equation}
where $f$ is the ion oscillator strength, $\lambda$ is the transition wavelength expressed in \AA, and $N$ is the ion column density in cm$^{-2}$\,(\kms)$^{-1}$. Combining equations~\ref{eq:fcov} and \ref{eq:N} yields an expression for $f_c$ as a function of $I$ and $N$. In cases where two or more transitions are measured for the same ion, from the same ground state, with different values of $f \lambda$, it is possible to solve these equations for $N$ and $f_c$. In the following analysis we will treat all variables as functions of velocity, i.e., $f_c(v)$.

For the low ionization species of interest, our spectra cover three such transitions of Si {\sc ii} at 1260, 1304, and 1526 \AA\, which we use to measure the covering fraction as a function of velocity $f_c(v)$, for each galaxy. Si {\sc ii} $\lambda$1304 is only used in the velocity range $v \gtrsim -200$ \kms\, where it is not contaminated by O {\sc i} $\lambda$1302. We resample the spectrum of each transition to a common velocity scale, and find the values of $N$ and $f_c$ which minimize the least-square residual $\chi^2 = \sum (I_{obs} - I_{N,f_c})^2 / \sigma_{obs}^2$ in each velocity bin. We additionally find the range of $N$ and $f_c$ for which $\chi^2$ is within 1 of the minimum value, and adopt this as the 1$\sigma$ uncertainty.

The best-fit $f_c$ and uncertainty calculated for each arc are shown as a function of velocity in Figure~\ref{fig:fcov}. Since Si {\sc ii} is the dominant ion of silicon in H {\sc i} gas, this is approximately equal to the covering fraction of H {\sc i} (provided that it is enriched with Si) which impedes the escape of ionizing radiation.

\subsection{Average Low-Ionization Absorption Profile}

A simple and complementary alternative to the method outlined in Section~\ref{sec:fcov} is to estimate the covering fraction from saturated transitions. In cases where $\tau \gg 1$, 
Equation~\ref{eq:fcov} simplifies to
\begin{equation}\label{eq:profile}
f_c = 1 - I/I_0.
\end{equation}
Several of the strongest absorption lines covered by our spectra are typically saturated, including: Si {\sc ii} $\lambda$1260, O~{\sc i} $\lambda$1302, Si {\sc ii} $\lambda$1304, C {\sc ii} $\lambda$1334, and Si {\sc ii} $\lambda$1526 which are all tracers of H {\sc i} gas. In order to minimize the statistical uncertainty we calculate the average intensity of these transitions as a function of velocity using an inverse-variance weighted mean, taking care not to use the wavelength region where O {\sc i} $\lambda$1302 and Si {\sc ii} $\lambda$1304 are blended (roughly $-300 \lesssim v \lesssim -200$ \kms\, depending on the kinematics of each source). These profiles are shown in Figure~\ref{fig:fcov} together with the covering fraction measured from Si {\sc ii}. We note that the covering fraction derived from Equation~\ref{eq:profile} is a strict lower limit.

\section{Results}\label{sec:discussion}

This work was motivated in large part by the need to disentangle kinematics and covering fractions of absorbing gas. In particular, we seek to explain the extent to which decreased absorption line equivalent widths measured from composite spectra in our earlier work \citep{jones2012} result from changes in gas kinematics compared to covering fractions, and the implications of this result for the escape of ionizing radiation. We are limited in examining the redshift evolution of these properties by the small number of sources with suitable spectra, and we caution that this sample is not necessarily representative of the LBG population at these redshifts. Nonetheless we can examine general trends within the existing data from this work and others in the literature \citep{pettini2002,quider2009,quider2010,dessauges-zavadsky2010}. 
Following the methods in Section~\ref{sec:analysis} we show the average absorption profiles and Si {\sc ii} covering fractions (derived from unblended transitions at 1260, 1526, and 1808 \AA) of well-studied $z=2$--3 galaxies for comparison in Figure~\ref{fig:fcov}. To quantify trends in the absorption line profiles, the velocity extent and maximum absorption depth for each galaxy are shown as a function of redshift in Figure~\ref{fig:abs}.

In the case of Abell 2390, we take the most conservative approach noting that the interpretation of the absorbing gas is complicated by the physical proximity and similar redshift of the two arcs. Their projected separation is $\sim 70$ kpc \citep{pello1999} and it is unclear which source lies in the foreground. At lower redshifts, \citet{steidel2010} have shown that low-ionization absorption seen in a background source at $b = 70$ kpc has a detectable average equivalent width $\sim 0.4$ \AA\ for the transitions of interest. Since we lack specific information about the 3-D geometry, the following analysis does not include any contribution from this effect. If anything, our results will overestimate the true covering fraction and therefore yield a more conservative constraint on the escape fraction.

\subsection{Kinematics}

\begin{figure*}
\begin{center}
\includegraphics[width=0.85\textwidth]{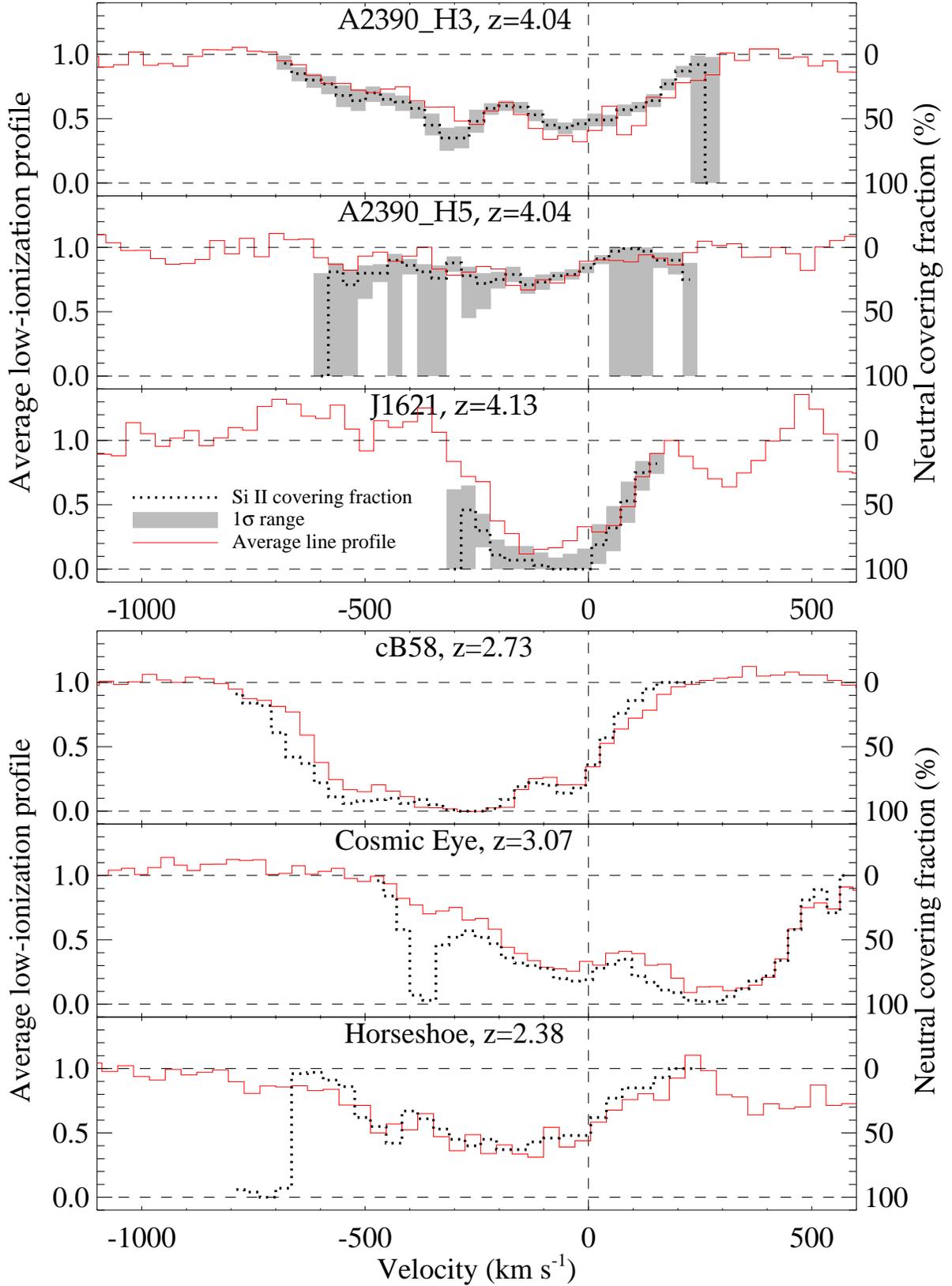}
\caption{
\label{fig:fcov}
Mean low-ionization absorption line profile and its associated neutral gas covering fraction derived using the methods discussed in Section~\ref{sec:analysis}. Si {\sc ii} covering fraction of the $z\simeq4$ arcs is measured from smoothed spectra with FWHM resolution $\simeq 110$ \kms, while average line profiles are from unsmoothed data (FWHM $\simeq 70$ \kms). There is no significant difference in results derived from the smoothed and unsmoothed spectra. Velocities are relative to adopted systemic redshifts (Section~\ref{sec:redshift}), derived from absorption line centroids for A2390\_H3 and nebular emission in the other two cases. Equivalent measurements from ESI spectra of $z=2$--3 arcs are shown for comparison. Details of these ESI spectra can be found in \citet{pettini2002} and \citet{quider2009,quider2010}.
}
\end{center}
\end{figure*}

The kinematics of foreground low-ionization gas are revealed in the absorption line profiles shown in Figure~\ref{fig:fcov}. In all cases we see significant blueshifted absorption indicating outflows, as expected given the high star formation surface densities \citep{heckman2002}. The maximum outflow velocity at which absorption is detected is $-700$ \kms\, in A2390\_H3, with an uncertainty of $\sim125$ \kms\, since we do not directly measure the systemic redshift. A2390\_H5 reveals weak absorption extending to $-600$ \kms, seen also in higher-ionization Si {\sc iv} and C {\sc iv} lines, although it is only marginally detected at $<-300$ \kms. The maximum outflow velocity in J1621 is $-300$ \kms.

The outflowing low-ionization gas attains a somewhat lower ($\sim30$\% on average) maximum velocity at higher redshift. This trend is not due to lower quality data as it remains evident in Figure~\ref{fig:fcov} if we consider alternative measures such as the FWHM or an absorption threshold at 25\% of the continuum flux. However all galaxies except J1621 have similar maximum velocities ranging from $600-800$ \kms. Likewise the extent of redshifted absorption is approximately $+200$ \kms\ for all sources, with the notable exception of the Cosmic Eye as discussed in detail by \citet{quider2010}; this indicates little difference in line broadening from rotation or other internal kinematic structure. Therefore, in this limited sample, the low-ionization gas kinematics are similar with a somewhat lower average velocity extent at higher redshift.

\subsection{Covering Fraction}

The covering fraction of each galaxy as a function of gas velocity is estimated from the methods described in Section~\ref{sec:analysis} and shown in Figure~\ref{fig:fcov}. Both methods are generally in good agreement indicating that Equation~\ref{eq:profile} is a valid approximation. 
Si {\sc ii} covering fractions derived for the Cosmic Eye are systematically higher than indicated by the average absorption profile; this is largely an artifact caused by additional absorption at the wavelength of Si {\sc ii} $\lambda$1260 from intervening gas at $z=2.66$ \citep{quider2010}. It is also apparent from Figure~\ref{fig:fcov} that the Si {\sc ii} covering fraction is poorly constrained in regions of weak absorption due to the marginal significance of individual absorption lines. This is most problematic in the high-velocity wings. The strong anticorrelation between absorption line strength and covering fraction, as well as results at lower redshift \citep{martin2009}, suggest that the most likely solution for such ambiguous cases is a low covering fraction of optically thick gas. We therefore opt to compare galaxies on the basis of their average absorption line profile as this quantity is simpler to define and less susceptible to the uncertainties described above. Nonetheless the Si {\sc ii} results are an important verification that the average profile accurately traces $f_c$.

\begin{figure}
\begin{center}
\includegraphics[width=0.5\textwidth]{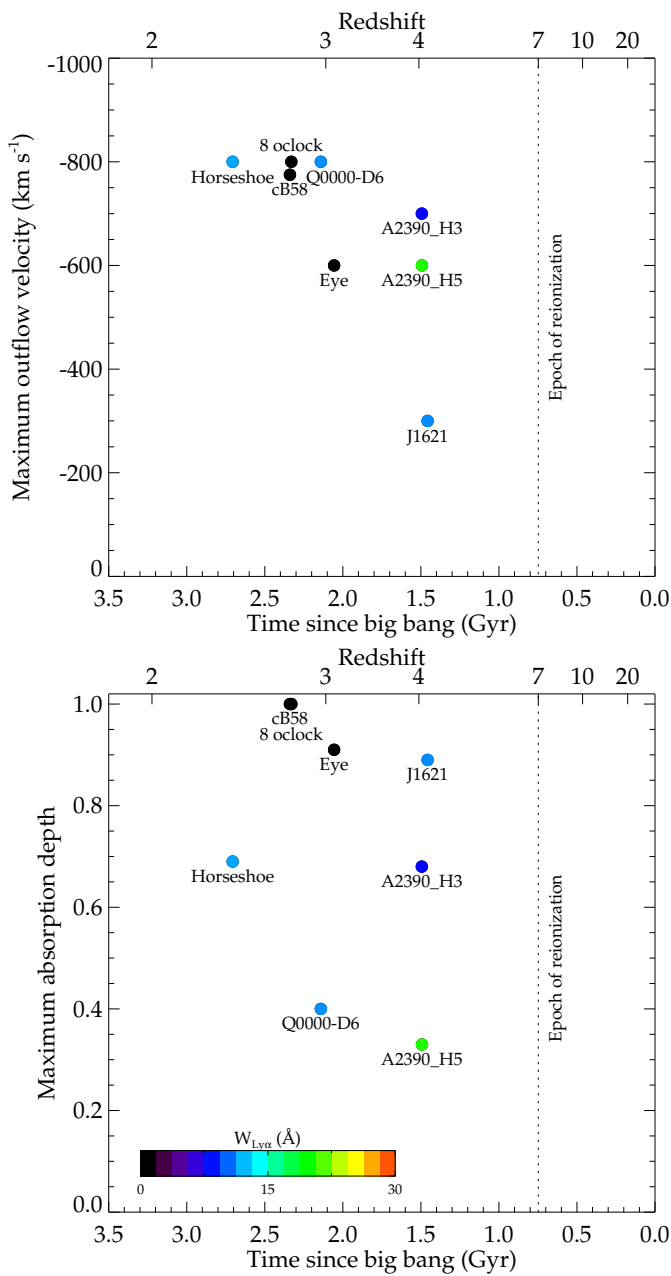}
\caption{
\label{fig:abs}
Outflow kinematics and covering fraction as a function of redshift. Top: maximum blueshifted velocity at which outflows are detected. In the case of A2390\_H3 there is an uncertainty of $\sim125$ \kms\ arising from the unknown systemic redshift. Bottom: maximum depth of the average absorption profiles shown in Figure~\ref{fig:fcov}, which serves as a proxy for the covering fraction of H {\sc i} at the corresponding velocity. 
We include estimated values for the 8 o'clock arc \citep{dessauges-zavadsky2010} and a somewhat lower resolution (R=1300) spectrum of Q0000-D6 \citep{steidel2010} in addition to the sources shown in Figure~\ref{fig:fcov}. Galaxies are color-coded according to their $W_{\rm Ly\alpha}$ and relevant data are listed in Table~\ref{tab:spec}.
On average, the $z=4$ galaxies show somewhat lower maximum outflow velocities and lower covering fractions (weaker absorption depth) compared to similarly studied sources at lower redshift, but the strongest trend is a decreasing covering fraction with $W_{\rm Ly\alpha}$. The $z=4$ data enable us to examine trends of H {\sc i} covering fraction at a time significantly closer to the epoch of reionization, thought to end at $z\simeq7$ (e.g., \citealt{schenker2012}).
}
\end{center}
\end{figure}

We can now compare the covering fractions measured at $z=4$ with sources at lower redshift. Figure~\ref{fig:abs} shows the maximum absorption depth for each galaxy as a function of redshift. The $z=4$ galaxies have maximum absorption depths corresponding to $f_c=0.3$--0.9, in each case occurring at $v\sim-100$ \kms. There is a large scatter in Figure~\ref{fig:abs} with $\sigma(f_{c,max}) = 0.26$ and no strong redshift dependence. Galaxies at $z=4$ do, however, have covering fractions which are lower on average by 25\% or $\Delta f_{c,max} = 0.16$ compared to $z=2$--3.

\subsection{Trends with \Lya}

We now turn to trends with \Lya\ equivalent width. Previous sections focused on possible redshift evolution of low-ionization absorption lines in our quest to examine whether this may signify an increasing ionizing escape fraction. The connection with \Lya\ is a natural one to explore given there is a strong correlation between its equivalent width $W_{\rm Ly\alpha}$ and low-ionization absorption \citep{jones2012,shapley2003}. Since our previous work has suggested that the distribution of $W_{\rm Ly\alpha}$ for LBGs of a fixed luminosity increases with redshift \citep{stark2010,stark2011,schenker2012}, we can hope to derive inferences about the low ionization absorption in sources for which \Lya\ measurements are now widely available.

\begin{figure}
\begin{center}
\includegraphics[width=0.5\textwidth]{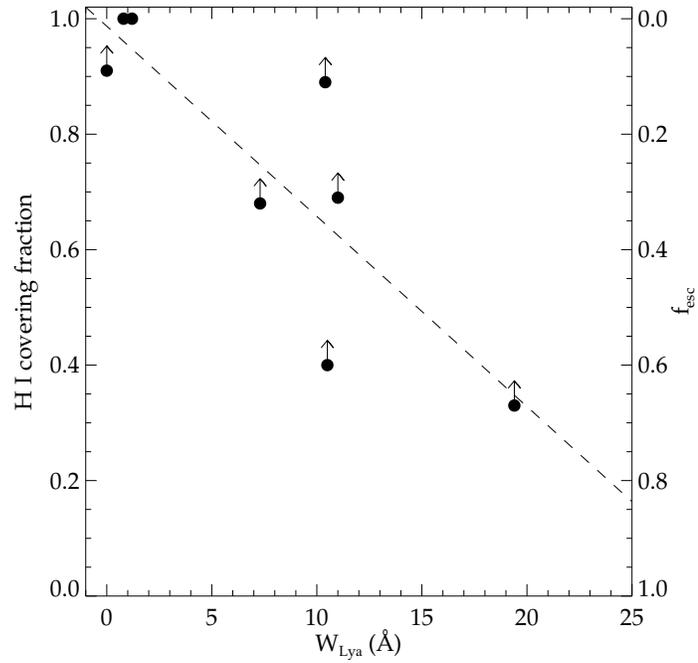}
\caption{
\label{fig:lya}
Correlation of \Lya\ equivalent width with H {\sc i} covering fraction as traced by low-ionization absorption lines. The dashed line shows a linear fit to the data. Here $W_{\rm Ly\alpha}$ includes only the emission component, and the two points at (1,1) are slightly offset for clarity. As discussed in the text, these absorption line measurements correspond to a lower limit on $f_c$ and an upper limit on $f_{esc}$. Since \Lya\ emission strength increases with redshift \citep{stark2010,stark2011}, this result likely indicates lower covering fractions (which would permit higher escape fractions) at earlier times.
}
\end{center}
\end{figure}

Each galaxy in Figure~\ref{fig:abs} is color-coded according to $W_{\rm Ly\alpha}$. This value refers only to the equivalent width of \Lya\ emission, differing from the conventional net sum of emission and absorption. The maximum outflow velocity is lower on average in galaxies with stronger \Lya\ emission, consistent with well-quantified results from composite spectra \citep{shapley2003}. More interestingly, Figure~\ref{fig:abs} reveals a trend of lower absorption depth (implying lower $f_c$) with stronger \Lya\ emission at $3.5\sigma$ significance. We show this relation in Figure~\ref{fig:lya}. In contrast, the trend of lower average $f_c$ at higher redshift has limited significance ($1.1\sigma$) and is explained by \Lya\ demographics within the sample. Although limited by the small sample size, this is an important first quantitative result at these redshifts. 
Since the frequency and equivalent width of \Lya\ emission increases in LBGs at higher $z=3\rightarrow6$, these results imply that the average covering fraction of low-ionization gas should decrease with redshift.

\subsection{Ionizing Escape Fraction}

Direct measurements of the ionizing flux are impractical at the redshifts of interest in this paper both because of the faint apparent magnitudes of LBGs and the high opacity of the \Lya\ forest. Nonetheless we can provide important constraints on $f_{esc}$ using indirect tracers of H {\sc i}.

Before doing so, we consider the potential systematic uncertainties which may limit our ability to estimate the true value of $f_{esc}$ from metal absorption lines. Results at $z\simeq3$ have shown that $f_{esc}$ is indeed dependent on low-ionization absorption strength \citep{bogosavljevic2010} although this relation is not one-to-one, likely due to the factors described below. While in general these preclude accurate estimates of the escape fraction, the maximum absorption depth (Figures~\ref{fig:fcov}, \ref{fig:abs}) is a valuable constraint on H {\sc i} spatial homogeneity and sets a stringent upper limit on $f_{esc}$.

1. We measure covering fraction as a function of velocity, yet gas at different velocities may cover different spatial regions and we lack the spatial resolution needed to evaluate this effect. The maximum absorption depth {\em at a given velocity} is thus a lower limit on the total H {\sc i} covering fraction and an upper limit on $f_{esc}$.

2. Metal-free H {\sc i} is not detected. To date only two instances of metal-free gas have been found at these redshifts \citep{fumagalli2011} and so this is likely insignificant. Again, this possibility implies that covering fractions measured in Section~\ref{sec:analysis} yield upper limits on $f_{esc}$.

3. Low column density gas will not be detected, although such gas will not affect $f_{esc}$ unless it has very low metallicity $\lesssim 0.1$ Z$_{\odot}$. The optical depth of metal transitions used here compared to Lyman continuum at $\sim900$ \AA\ is $\frac{\tau}{\tau_{\rm LyC}} = 1.7$ for the weakest line (Si {\sc ii} $\lambda$1304) and 10--20 for the strongest transitions (Si {\sc ii} $\lambda$1260, C {\sc ii} $\lambda$1334, O {\sc i} $\lambda$1302) for solar abundance ratios \citep{asplund2009}. Galaxies in the $z=2$--3 sample with measured interstellar abundance ratios have Z $\gtrsim0.4$ Z$_{\odot}$ for the relevant elements, such that the attenuation of weaker Si lines is roughly equal to that of ionizing radiation.

4. There may be narrow components with uniform covering fraction which are spectroscopically unresolved. However, such smooth absorption and column density profiles would require a remarkably regular spacing of discrete narrow components, and so a partial covering fraction appears more likely (see \citealt{pettini2002,quider2009}).

5. Si {\sc ii} and C {\sc ii} are present in both H {\sc i} and H {\sc ii} regions. We have therefore confirmed that O {\sc i}, with essentially the same ionization potential as H {\sc i}, gives consistent results in all cases.

6. We measure a covering fraction at 1260--1526 \AA, but the stars which are bright at these wavelengths do not necessarily emit at $<$912 \AA. Our measurements therefore correspond to a constraint on the relative escape fraction $f_{esc} = L_{\rm LyC} / L_{1500}$ as commonly used in the literature. However, the spatial distribution of $L_{1500}$ is similar to ionizing emission as traced by Balmer lines in $z\simeq1-3$ galaxies (e.g., \citealt{jones2010}), indicating that this effect may be minimal.

With these caveats in mind, we now summarize what can be learned about the ionizing escape fraction at $z=4$. The depth of low-ionization absorption lines gives upper limits $f_{esc} < 0.7$ for A2390\_H5, $<0.3$ for A2390\_H3, and $<0.1$ for J1621. The true values are likely well below the upper limits and in three lower-redshift galaxies we can verify that this is the case. \citet{quider2010} show that the Cosmic Eye has an H {\sc i} covering fraction of $\simeq 95$\% implying $f_{esc} < 0.05$ based on damped \Lya\ absorption, more stringent than the $f_{esc} < 0.1$ from our analysis. Secondly, no ionizing radiation is detected from the Horseshoe in deep UV imaging (B. Siana, private communication) nor from the spectrum of Q0000-D6 ($f_{esc} < 0.16$; \citealt{giallongo2002}) despite low maximum covering fractions (Figures~\ref{fig:fcov}, \ref{fig:abs}). Non-uniform coverage of low-ionization metals is evidently a necessary, but not sufficient, condition for the escape of ionizing radiation. Results derived in this paper should therefore be strictly interpreted as upper limits. Nonetheless, our measurements at $z=4$ are readily compatible with the value $f_{esc} \gtrsim 0.2$ required for galaxies to reionize the universe \citep{robertson2013}.

\section{Discussion}

In our previous work \citep{jones2012} we found a decrease in the low-ionization absorption equivalent width with redshift in composite LBG spectra, but were unable to distinguish whether gas kinematics and/or covering fraction were the cause. The new data presented in this paper now resolve this ambiguity. Although the sample of LBGs with high quality spectra at $z>2$ is small and not necessarily representative,  the present data show an approximately equal decrease of $\sim25$\% in the gas velocity and covering fraction with redshift. The much stronger dependence of low-ionization absorption with \Lya\ emission appears to be predominantly due to the covering fraction of neutral hydrogen.

The main limitation of methods used in this work to constrain $f_{esc}$ is that we do not directly measure the fraction of ionizing emission covered by H {\sc i}. While the spectrally-resolved covering fraction of heavy elements provides an important measure of the ``patchiness'' of H {\sc i}, gas at different velocities may cover different spatial regions and so we can derive only a lower (upper) limit on $f_c$ ($f_{esc}$). Deep integral field spectroscopy with good spatial resolution may resolve this issue and provide better constraints on $f_c$ and $f_{esc}$.

We have successfully measured the kinematics and covering fraction of low-ionization (H~{\sc i}) gas in $z=4$ LBGs from high quality rest-UV spectra. Resulting upper limits on the ionizing escape fraction $f_{esc} \leq 1 - f_c$ are readily consistent with that required for star-forming galaxies to reionize the universe (e.g., \citealt{robertson2013}). Importantly, the new data enable these measurements at a time $< 1$ Gyr from the reionization epoch, much earlier than was previously possible. We note that in order for galaxies to reionize the universe, the escape fraction must increase rapidly from $\left< f_{esc} \right> = 0.05$ measured at $z=3$ \citep{bogosavljevic2010} to a value $f_{esc} \gtrsim 0.2$ at $z=7$ \citep{robertson2013,kuhlen2012}.  While the trends with redshift are poorly constrained, the available data reveal a reduced covering fraction with increasing $W_{\rm Ly\alpha}$ indicating that galaxies with moderate or strong \Lya\ emission are likely to have larger $f_{esc}$. This is supported by direct evidence at $z=3$, where galaxies with detectable ionizing flux have stronger \Lya\ emission and weaker low-ionization absorption than LBGs with lower $f_{esc}$ \citep{bogosavljevic2010}. Clearly the distribution of (metal-enriched) H {\sc i} is patchier in galaxies with stronger \Lya\ emission, and therefore more likely that ionizing radiation can escape.  Since the frequency and strength of \Lya\ emission in typical LBGs increases with redshift \citep{stark2010,stark2011}, our results provide new evidence that the covering fractions decrease (and therefore $f_{esc}$ increases) with redshift.

\acknowledgements

We thank Max Pettini for providing ESI spectra of the $z=2$--3 galaxies used as a comparison sample.
T.A.J. acknowledges support from the Southern California Center for Galaxy Evolution through a CGE Fellowship.
D.P.S. acknowledges support from NASA through Hubble Fellowship grant \#HST-HF-51299.01 awarded by the Space Telescope Science Institute, which is operated by the Association of Universities for Research in Astronomy, Inc, for NASA under contract NAS5-265555. 
The analysis pipeline used to reduce the DEIMOS data was developed at UC Berkeley with support from NSF grant AST-0071048.
This work relies on data obtained at the W.M. Keck Observatory, which is operated as a scientific partnership among the California Institute of Technology, the University of California and the National Aeronautics and Space Administration, and was made possible by the generous financial support of the W.M. Keck Foundation.
We wish to recognize the significant cultural role that the summit of Mauna Kea has within the indigenous Hawaiian community; we are most fortunate to have the opportunity to conduct observations from this mountain.

\begin{deluxetable*}{lcccccccl}
\tablecolumns{8}
\tablewidth{0pt}
\tablecaption{\bf Gravitationally Lensed Sources\label{tab:arcs}}

\tablehead{
\colhead{ID} & \colhead{RA} & \colhead{Dec}
& \colhead{$z$} & \colhead{$m_{\rm AB}$} & \colhead{$M_{\rm UV}$\tablenotemark{a}} & \colhead{t$_{\rm exp}$ (hrs)} & \colhead{References}}

%\multicolumn{8}{c}{ID}&{RA}&{Dec}&{Redshift}&{$z'_{AB}$}&{$M_{UV}$(unlensed)}&{Exposure(s)}&{References}
\medskip
\startdata
Abell 2390\_H3a & 21:53:34 & +17:42:03 & $4.043 \pm0.002$  & $I=22.4$\tablenotemark{b} & $-21.3\pm0.3$ & 10.8  & \citet{frye1998,pello1999} \\
Abell 2390\_H5b & 21:53:35 & +17:41:33 & $4.0448\pm0.0003$ & $I=22.8$\tablenotemark{b} & $-21.0\pm0.3$ & 10.8  & \citet{pello1999} \\
1621+0607       & 16:21:33 & +06:07:05 & $4.1278\pm0.0004$ & $I=22.6$\tablenotemark{c} & $-21.0$\tablenotemark{d}       & 5.7   & \citet{bayliss2011} \\
\smallskip \\
cB58  & 15:14:22 & +36:36:25 & 2.73 & $r=20.6$ & $-21.1$ & & \citet{pettini2002} \\
Cosmic Eye  & 21:35:13 & $-$01:01:43 & 3.07 & $r=20.3$ & $-21.8$ & & \citet{quider2010} \\
Horseshoe  & 11:48:33 & +19:29:59 & 2.38 & $g=20.1$ & $-21.5$ & & \citet{quider2009} \\
8 o'clock arc & 00:22:41 & +14:31:10 & 2.73 & $r=19.2$ & $-23.4^{+0.9}_{-0.4}$ & & \citet{dessauges-zavadsky2010} \\
Q0000-D6\tablenotemark{e} & 00:03:24 & $-$26:02:49 & 2.96 & $R=22.9$ & $-22.6$ & & \citet{steidel2010,giallongo2002} \\
%\hline
\tablenotetext{a}{Unlensed absolute magnitudes at effective wavelength $\sim 1500$ \AA\ depending on filter and redshift.}
\tablenotetext{b}{HST/WFPC2 F814W magnitude from \citet{sand2005}.}
\tablenotetext{c}{Derived from the spectrum at the mean wavelength of HST/WFPC2 F814W, $\lambda_{\rm mean} = 8369.5$ \AA.}
\tablenotetext{d}{For a magnification factor $\mu = 10$.}
\tablenotetext{e}{This source is not gravitationally lensed.}
\end{deluxetable*}

\begin{deluxetable*}{lcccccccc}
\tablecolumns{8}
\tablewidth{0pt}
\tablecaption{\bf Spectroscopic Properties\label{tab:spec}}

\tablehead{
\colhead{ID} & \colhead{$W_{\rm Ly\alpha}$\tablenotemark{a} (\AA)} & \colhead{$f_c$\tablenotemark{b}} & \colhead{$v_{max}$ (\kms)} 
}
\medskip
\startdata
Abell 2390\_H3 & 7.3          & 0.68      & 700 \\
Abell 2390\_H5 & 19.4         & 0.33      & 600 \\
SDSS J1621     & 10.4         & 0.89      & 300 \\
cB58           & $\sim$1      & 1.0       & 775 \\
Cosmic Eye     & $\sim$0      & 0.91      & 600 \\
Horseshoe      & 11           & 0.69      & 800 \\
8 o'clock arc  & $\sim$1      & $\sim$1.0 & 800 \\
Q0000-D6       & 10.5         & $\sim$0.4 & 800 \\
\tablenotetext{a}{Emission component only.}
\tablenotetext{b}{Lower limit derived from absorption line depth.}
\end{deluxetable*}

\medskip

\begin{thebibliography}{43}
\expandafter\ifx\csname natexlab\endcsname\relax\def\natexlab#1{#1}\fi

\bibitem[Asplund et al.(2009)]{asplund2009}
Asplund, M., Grevesse, N., Sauval, A.~J., \& Scott, P.\ 2009, \araa, 47, 481 

\bibitem[{{Bayliss} {et~al.}(2011)}]{bayliss2011}
{Bayliss}, M.B., {Hennawi}, J.F., {Gladders}, M.D. {et~al.} 2011, \apjs, 193, 8

\bibitem[Bernardi et al.(2003)]{bernardi2003}
Bernardi, M., et al.\ 2003, \aj, 125, 32 

\bibitem[Bogosavljevi{\'c}(2010)]{bogosavljevic2010}
Bogosavljevi{\'c}, M.\ 2010, Ph.D.~Thesis

\bibitem[Dessauges-Zavadsky et al.(2010)]{dessauges-zavadsky2010} 
Dessauges-Zavadsky, M., D'Odorico, S., Schaerer, D., Modigliani, A., Tapken, C., \& Vernet, J.\ 2010, \aap, 510, A26 

\bibitem[Ellis et al.(2013)]{ellis2013}
Ellis, R.~S., et al.\ 2013, \apjl, 763, L7 

\bibitem[Erb et al.(2010)]{erb2010}
Erb, D.~K., Pettini, M., Shapley, A.~E., Steidel, C.~C., Law, D.~R., \& Reddy, N.~A.\ 2010, \apj, 719, 1168 

\bibitem[{{Fosbury} {et~al.}(2003)}]{fosbury2003}
{Fosbury}, R.A.E.,  {Villar-Mart\'in}, M., .{Humphrey}, A. {et~al.}, 2003, \apj, 596, 797

\bibitem[{{Frye} \& {Broadhurst}(1998)}]{frye1998}
{Frye}, B. \& {Broadhurst}, T.J. 1998 \apj, 499, L115

\bibitem[Fumagalli et al.(2011)]{fumagalli2011}
Fumagalli, M., O'Meara, J.~M., \& Prochaska, J.~X.\ 2011, Science, 334, 1245 

\bibitem[Giallongo et al.(2002)]{giallongo2002}
Giallongo, E., Cristiani, S., D'Odorico, S., \& Fontana, A.\ 2002, \apjl, 568, L9 

\bibitem[{{Giavalisco} {et~al.}(2004)}]{giavalisco2004}
{Giavalisco}, M., {Ferguson}, H., {Koekemoer}, A. {et~al.}, 2004, \apj, 600, 93

\bibitem[Heckman(2002)]{heckman2002}
Heckman, T.~M.\ 2002, Extragalactic Gas at Low Redshift, 254, 292 

\bibitem[Hinshaw et al.(2012)]{hinshaw2012}
Hinshaw, G., et al.\ 2012, arXiv:1212.5226 

\bibitem[Jones et al.(2010)]{jones2010}
Jones, T.~A., Swinbank, A.~M., Ellis, R.~S., Richard, J., \& Stark, D.~P.\ 2010, \mnras, 404, 1247 

\bibitem[{{Jones} {et~al.}(2012){Jones}, {Stark} \& {Ellis}}]{jones2012}
{Jones}, T.~A., {Stark}, D.~P., \& {Ellis}, R.~S. 2012, \apj, 751, 51

\bibitem[Kuhlen \& Faucher-Gigu{\`e}re(2012)]{kuhlen2012}
Kuhlen, M., \& Faucher-Gigu{\`e}re, C.-A.\ 2012, \mnras, 423, 862 

\bibitem[Martin \& Bouch{\'e}(2009)]{martin2009} Martin, C.~L., \& Bouch{\'e}, N.\ 2009, \apj, 703, 1394

\bibitem[Mortlock et al.(2011)]{mortlock2011}
Mortlock, D.~J., et al.\ 2011, \nat, 474, 616 

\bibitem[{{Nestor} {et~al.}(2012)}]{nestor2012}
{Nestor}, D.~B., {Shapley}, A.~E., {Kornei}, K.~A. {et~al.}, 2012, arXiv 1210:2393

\bibitem[Oesch et al.(2013)]{oesch2013}
Oesch, P.~A., et al.\ 2013, arXiv:1301.6162 

\bibitem[{{Oke}(1974){Oke}}]{oke1974}
{Oke}, J.~B., 1974 \apjs, 27, 21

\bibitem[Pell{\'o} et al.(1999)]{pello1999} Pell{\'o}, R., et al.\ 1999, \aap, 346, 359 

\bibitem[{{Pettini} {et~al.}(2002)}]{pettini2002}
{Pettini}, M., {Rix}, S.A., {Steidel}, C.C. {et~al.} 2002, \apj, 569, 742

\bibitem[{{Quider} {et~al.}(2009)}]{quider2009}
{Quider}, A.M., {Pettini}, M., {Shapley}, A.E. \& {Steidel}, C.C. 2009, \mnras, 398, 1263

\bibitem[{{Quider} {et~al.}(2010)}]{quider2010}
{Quider}, A.M., {Shapley}, A.E., {Pettini}, M., {Steidel}, C.C. \& {Stark}, D.P. 2010, \mnras, 402, 1467

\bibitem[{{Robertson} {et~al.}(2010){Robertson}, {Ellis}, {Dunlop}, {McLure},
  \& {Stark}}]{robertson2010}
{Robertson}, B.~E., {Ellis}, R.~S., {Dunlop}, J.~S., {McLure}, R.~J., \& {Stark}, D.~P. 2010, \nat, 468, 49

\bibitem[Robertson et al.(2013)]{robertson2013}
Robertson, B.~E., et al.\ 2013, arXiv:1301.1228 

\bibitem[{{Sand} {et~al.}(2005)}]{sand2005} Sand, D.~J., Treu, T., Ellis, R.~S., \& Smith, G.~P.\ 2005, \apj, 627, 32 

\bibitem[{{Schenker} {et~al.}(2012)}]{schenker2012}
{Schenker}, M.~A., {Stark}, D.~P., {Ellis},  R.~S. {\it et al.}. 2012, \apj, 744, 179

\bibitem[{{Shapley}(2011){Shapley}}]{shapley2011}
{Shapley}, A.~E. 2011 \araa, 49, 525

\bibitem[{{Shapley} {et~al.}(2003){Shapley}, {Steidel}, {Pettini}, \&
  {Adelberger}}]{shapley2003}
{Shapley}, A.~E., {Steidel}, C.~C., {Pettini}, M., \& {Adelberger}, K.~L. 2003, \apj, 588, 65

\bibitem[{{Stark} {et~al.}(2009)}]{stark2009}
{Stark}, D.~P., {Ellis}, R.~S., {Bunker}, A. {et~al.} 2009, \apj, 697, 1943

\bibitem[{{Stark} {et~al.}(2010)}]{stark2010}
{Stark}, D.~P., {Ellis}, R.~S., {Chiu}, K., {Ouchi}, M. \& {Bunker}, A. 2010, \mnras, 408, 1628

\bibitem[{{Stark} {et~al.}(2011)}]{stark2011}
{Stark}, D.~P., {Ellis}, R.~S., \& {Ouchi}, M. 2011, \apj, 728, L2

\bibitem[Steidel et al.(2001)]{steidel2001}
Steidel, C.~C., Pettini, M., \& Adelberger, K.~L.\ 2001, \apj, 546, 665 

\bibitem[{{Steidel} {et~al.}(2010)}]{steidel2010}
{Steidel}, C.~C., {Erb}, D.K., {Shapley}, A.E. {et~al.}, 2010, \apj, 717, 289

\bibitem[{{Steidel} {et~al.}(2011)}]{steidel2011} Steidel, C.~C., Bogosavljevi{\'c}, M., Shapley, A.~E., Kollmeier, J.~A., Reddy, N.~A., Erb, D.~K., \& Pettini, M.\ 2011, \apj, 736, 160 

\bibitem[{{Vanzella} {et~al.}(2008)}]{vanzella2008}
{Vanzella}, E., {Cristiani}, S., {Dickinson}, M.E. {et~al.}, 2008, \aa, 478, 83

\bibitem[Weiner et al.(2009)]{weiner2009}
Weiner, B.~J., et al.\ 2009, \apj, 692, 187 

\end{thebibliography}
\end{document}